# A platform for analysis of nanoscale liquids with an integrated sensor array based on 2-d material

Michael Engel[1,2], Peter W. Bryant[1], Rodrigo F. Neumann[1], Ronaldo Giro[1], Claudius Feger[2], Phaedon Avouris[2], Mathias Steiner[1,2,*]

[1] *IBM Research, Rio de Janeiro, RJ 22290-240, Brazil*

[2] *IBM Research, Yorktown Heights, NY 10598, USA*

[*]*msteine@us.ibm.com, mathiast@br.ibm.com*

**ABSTRACT:** Analysis of nanoscale liquids, including wetting and flow phenomena, is a scientific challenge with far reaching implications for industrial technologies. We report the conception, development, and application of an integrated platform for the experimental characterization of liquids at the nanometer scale. The platform combines the sensing functionalities of an integrated, two-dimensional electronic device array with *in situ* application of highly sensitive optical micro-spectroscopy and atomic force microscopy. We demonstrate the performance capabilities of the platform with an embodiment based on an array of integrated, optically transparent graphene sensors. The application of electronic and optical sensing in the platform allows for differentiating between liquids electronically, for determining a liquid's molecular fingerprint, and for monitoring surface wetting dynamics in real time. In order to explore the platform's sensitivity limits, we record topographies and optical spectra of individual, spatially isolated sessile oil emulsion droplets having volumes of less than ten attoliters. The results demonstrate that integrated measurement functionalities based on two-dimensional materials have the potential to push lab-on-chip based analysis from the microscale to the nanoscale.

**MAIN TEXT**

Two-dimensional materials are currently under investigation for optical and electronic sensing applications [1]. The incorporation of two-dimensional materials could enable novel sensor conceptions for the analysis of liquids at the nanometer scale, e.g. [2], if scalable integration within dedicated device platforms can be achieved. Ideally, such a platform should allow the simultaneous interrogation of optical, electronic, and mechanical parameters that are needed for a proper physical and chemical characterization of the system under study. Graphene [3], due to its combination of optical transparency, electric conductivity, mechanical flexibility, and chemical inertness is a promising functional sensor material [4], specifically in lab-on-chip and microfluidics applications [5,6,7,8]. As recent demonstrations have shown [9,10,11], the same properties are also highly beneficial in the context of biomedical applications [12]. For quantitative physical and chemical analysis of liquids at microscopic scales, however, various complementary measurements have to be performed *in situ*. Given the stringent requirements imposed by nanometer scale experiments, a dedicated lab-on-chip platform with integrated 2-d liquid sensing functionalities is needed that enables simultaneous application of high-resolution physical and chemical characterization techniques. In this Letter, we introduce a liquid analysis platform with integrated 2-d material based sensor array for electric, optical, and topographical characterization of liquids at the nanometer scale. We demonstrate that the platform allows to differentiate between liquids by electronic means and, based on its optical transparency, enables a spectroscopic assessment of a liquid's chemical identity. In addition, by correlating optical and electronic signals as functions of time, we track wetting dynamics at the platform surface with millisecond resolution. Finally, we validate the platform's ultimate sensitivity by mapping *in situ* the surface contact area and topography of a spatially isolated, nanometer scale oil emulsion droplet and by recording the droplet's fluorescence spectrum. Potential applications of the 2-d material-based sensing platform include the quantitative analysis of surface wetting and flow phenomena and (bio-) chemical reactions at small length scale.

A visual representation of the analysis platform with various measurement modes is shown in Fig.1a. The photograph in Fig.1b shows the optically transparent platform having dimensions that are adapted for the combined application of optical micro-spectroscopy and atomic force microscopy. As a representative 2-d material for the platform embodiment we have chosen graphene. The integrated device array located at the platform center comprises 400 functional sites that are optimized for combined electronic and optical sensing, see Fig.1b,c.

For optical and electronic measurements at a specific device site as shown in Fig.2a, the platform is integrated with an inverted optical microscope equipped with an immersion objective as schematically indicated in Fig.2b. A dedicated electric probe system allows to address individual devices in the array and to connect them to an external measurement and control system. For optical measurements, the platform can be raster scanned with respect to the microscope objective for confocal image formation and point-wise spectroscopy. Alternatively, wide field microscopy images and atomic force microscopy images can be obtained from the top of the platform. Also shown in Fig.2b is a cross sectional view of the integrated device stack, together with the electronic measurement scheme. Details regarding the manufacturing of the device platform are provided in the Methods Section.

In a first series of experiments, schematically indicated at the top of Fig.3a, we analyze how the deposition of liquids at the platform surface will affect the electric transport characteristics of an integrated 2-d sensor. For the electronic detection of liquids, the integrated graphene device is operated in a transistor configuration where the ITO layer functions as back gate electrode, see Fig.2. We note that in previous, graphene-based electronic sensor demonstrations liquid gating techniques were used instead [13,14]. As representative liquids we have chosen decane and water as they are visually indistinguishable, however, with different chemical properties to be identified within the platform. In Fig.3b, we record as a reference the electric transfer characteristics of the graphene device under ambient conditions, i.e. without liquid. The gate voltage sweep reveals that the integrated graphene layer is heavily p-doped, as expected, and it is not possible to identify the charge neutrality point of the device within the experimental gate voltage range. As we deposit decane at the platform surface, i.e. onto the $SiO_2$ layer capping the integrated graphene device, we observe that the electric device characteristics changes significantly. The overall electric current level is lower and the profile of the transfer characteristics is almost flat. In the case water is dispensed onto the device instead of decane, the electric current level is further reduced while the charge neutrality point occurs within the gate voltage range, close to $V_g$=2.5V. The results show that the integrated, graphene-based sensor is capable of differentiating between decane and water electronically through a dielectric cap layer with 20nm thickness. By repetition on the same device we have confirmed that the effect of liquid deposition on electric transport is reversible and reproducible and we observe the electric effect even for layer thickness of 120nm. The cap design has the benefit of protecting the graphene layer so that the platform can be cleaned and reutilized.

While we observe that the deposition of decane, a liquid with low polarity, reduces the overall gating efficiency in the device, the deposition of water, a liquid with much higher polarity, causes a significant charge carrier redistribution which is evidenced by the occurrence of graphene's charge neutrality point in the transfer characteristics of the device. The observed phenomenology can be rationalized by assuming that the capacitance at the graphene/$SiO_2$ interface is modulated due to carrier redistribution that occurs upon deposition of a liquid on top of the $SiO_2$-layer capping the graphene device. The back gate effect controlled by external bias voltage is superimposed by a liquid-induced gating effect with a coupling strength depending mainly on the liquid's polarity as well as the dielectric properties of the device stack. A potential application of the integrated sensor principle could be in all-electronic sensing of spatial variations in the composition of multiphase fluids in a lab-on-chip platform. Future research is needed in order to provide thickness dependent calibrations for relevant dielectrics integrated as cap layer. Also, a model of electric transport in the device is needed that quantifies the liquid-induced electrostatic gating effect in presence of the dielectric cap layer.

While different liquids cause different electric transfer characteristics in the graphene device, see Fig.3b, it is not straightforward to identify from electric transport modification the chemical composition of a liquid. In order to chemically identify the liquid under investigation, we perform Raman micro-spectroscopy within the active device area, at the liquid-$SiO_2$ interface, as indicated at the bottom of Fig.3a. The reference spectrum measured under ambient conditions, i.e. without liquid, exhibits the characteristic bands associated with Raman-active molecular vibrations in graphene, i.e. G and 2D [15]. In the case water is deposited at the platform surface, we observe in the Raman spectrum the characteristic band at 3400cm$^{-1}$ mainly due to the O-H stretch vibrations [16]. In the case of decane, the Raman spectrum exhibits characteristic fingerprint features between 1000cm$^{-1}$ and 1500cm$^{-1}$ and a strong band at about 2900cm$^{-1}$; see reference [17] for comparison. The results demonstrate that optical micro-spectroscopy and integrated electronic sensing can be performed to identify liquids within the analysis platform. Based on the knowledge of the chemical composition of the liquid, the electronic transport data can now be calibrated for specific sensing applications. It is conceivable that, alternatively, spectroscopic access to the platform could be provided by integrating optical waveguide structures [18], further increasing the platform's application potential.

The measurements in Fig.3 represent steady-state experimental conditions and the question remains if liquid deposition and surface wetting dynamics can be monitored in real time. In the following, we experimentally monitor the wetting dynamics upon droplet deposition in the time domain by correlating optical microscope imaging and electric transport measurements. In order to facilitate droplet deposition, we vaporize water with an ultra-sonication unit so that small water droplets ultimately reach the device site under study within a suitable time window. In order to improve clarity with regards to the measurement principle we visualize schematically the simultaneous optical and electronic measurement at the top of Fig.4a. The platform is optically imaged at a specific device location where the exposed graphene area has a size of 20 micron x 20 micron, see Fig.4b. By applying a bias voltage of $V_d$=0.1mV across the device we obtain an electric current of about $I_d$=120mA in the graphene layer forming the baseline for the measurement. A representative droplet deposition event is shown schematically in Fig.4a, from top to bottom, and experimentally in the optical image series Fig.4c-e. Within less than a second, a landing water droplet hits the platform surface, spreads across the entire device area, and contracts at the side of the device where it eventually evaporates; see position of the dark shadow in the optical images. Simultaneously, by recording the electric current through the biased graphene layer, we can monitor the strength of the liquid-solid interaction as function of time. The electric current through the graphene layer is flat for $t<0$s until a water droplet impacts the device surface at $t=0$s. This is evidenced by the faint shadow formation in the lower left corner close to the contact of the optical image in Fig.4c. The droplet touch down at the device surface causes an electric current peak having a width of about 250ms. After initially covering the entire device area, see image Fig.4d taken at $t=200$ms, the liquid forms a steady droplet shape at the lower right hand side of the device, see Fig.4e, where it slowly evaporates. As can be seen in Fig.4f, however, the liquid-induced modulation of the electric current through the graphene layer flattens out only after the liquid is completely evaporated. The differential electric signal in Fig.4g demonstrates that the baseline quality allows sensing the liquid-solid interaction with a signal-to-noise ratio of the order of 100. While in the present case the graphene layer is directly exposed to the liquid, we note that similar current modulations are also observed in case a capping $SiO_2$-layer is present. While these results demonstrate the potential of the graphene-based platform for studying the dynamics of surface wetting and liquid evaporation at small scales, future research will need to establish the spatial and temporal resolution limits that can be achieved with this 2d-material based, electronic sensing method.

A key question is to what extent the platform enables the analysis and characterization of nanometer scale liquids. In order to investigate resolution and sensitivity limits, we have deposited and characterized nanoscale droplets of an oil-surfactant emulsion at the platform surface by means of an ultra-sonication technique. In order to clarify the experimental arrangement, we schematically indicate in Fig.5a the application of confocal optical laser scanning microspectroscopy from the underside of the platform and the application of atomic force microscopy from the top. In Fig.5b, we show an atomic force microscope image that exhibit spatially isolated, individual nanoscale oil emulsion droplets with diameters as small as 100nm that reside at the device surface. The confocal elastic laser light scattering image of the same device area shown in Fig.5e exhibits dark features at the positions where oil droplets are in contact with the device surface. The optical contrast of oil emulsion droplets vanishes at droplet diameters of about 300nm, close to the optical resolution limit, as can be seen by comparing the optical signals in Fig.5e with the topographic signatures of the same oil droplets in Fig.5b. We have performed cross sectional image analysis, see Fig.5c,f, in order to estimate a droplet diameter of 350nm and a droplet height of 45nm for a spatially isolated and optically resolved oil emulsion droplet. A three-dimensional model fit based on Bessel function expansion to the droplet topography data of the same droplet is shown in Fig.5d. The fit provides a droplet volume of $(6.8\pm0.3)\cdot 10^{-18}$ liter. In order to acquire the fluorescence spectrum of this oil emulsion droplet we focus the laser onto the droplet center and spectrally analyze the scattered light with a spectroscopic unit. The measured fluorescence spectrum of the oil emulsion droplet, see Fig.5g, exhibits a broad emission band peaking at about 600nm with a full-width-at-half-maximum of about 75nm. While the above results highlight the potential of the platform for simultaneous electric, optical, and topographical characterization of liquids at nanoscale, further research is needed for investigating the sensitivity limits of the integrated electronic sensor with regards to liquid volume. For optimizing detection sensitivity, the use of 2-d semiconductors or their heterostructures within the platform could be a promising route.

Relevant areas for future application of the platform are the analysis of surface wetting and flow phenomena as well as chemical reaction analysis at the attoliter scale. To that end, the integrated 2-d material layer could provide various integrated functionalities, such as controlling the charge carrier density at the solid-liquid (wetting) interface, or acting as an integrated heater for controlling chemical reaction kinetics *in situ*. The simultaneous application of multiple measurement techniques in the platform to nanoscale liquid systems with higher complexity, e.g. multiple-phase liquids with nanoparticles or functional materials, will allow us to experimentally verify technologically relevant liquid-solid interactions for industrial scale applications.

In summary, we have introduced and characterized a measurement platform with an integrated 2-dimensional electric sensor array for the experimental investigation of liquids and liquid-solid interactions at the nanometer scale. The platform implementation with a graphene-based electronic sensor array allows monitoring liquid deposition and wetting dynamics in real time and to confirm a liquid's chemical identity. In order to explore sensitivity limits, we have measured the contact area, topography, and fluorescence spectrum of the same spatially isolated, nanoscale oil emulsion droplet with a volume of less than 10 attoliters located at the platform surface by combining optical micro-spectroscopy and atomic force microscopy. In conclusion, the analysis platform provides experimental access to and control of nanoscale liquids with potential future applications in industrial analytics ranging from chemical engineering to biotechnology and natural resources recovery. Integrated measurement functionalities based on 2-dimensional materials such as graphene have the potential to push lab-on-chip based analysis to the nanoscale.

**METHODS SECTION**

The device platform is built on optically transparent ITO-coated glass slides having a size of 20x20mm$^2$ and a thickness of 170 micron. The ITO thickness is about 200nm with a resistance of 8-12Ohm/sq. The ITO layer serves as a charge dissipation layer during e-beam lithography and as a back gate electrode for the device array. In a first step, we deposit 100nm of $Al_2O_3$ by a high temperature (250°C) atomic layer deposition (ALD) on top of the ITO layer. In a second step, we transfer CVD-grown graphene [19] by a PMMA assisted technique [20] onto the device stack. Next we perform e-beam lithography followed by metal evaporation (5nm Ti/50nm Au) and lift-off to define alignment marks and a first set of larger contact pads. We then define the active graphene device area by patterning a negative tone resist bilayer (PMMA/HSQ), followed by an oxygen plasma step to etch the exposed graphene. Subsequently, we define metallic contacts to the patterned graphene by e-beam lithography. This step is followed by metal evaporation (5nm Ti/50nm Au) and a lift-off step. In a final step, we cap the device with a 20nm (or 120nm) thick $SiO_2$ layer which is deposited by thermal evaporation. Depending on the specifics of the application, the surface of the dielectric cap layer could be further modified and functionalized [21]. Manufacturing and materials details regarding earlier versions and implementations of the optical transparent device stack can be found in references [22,23,24].

**ACKNOWLEDGEMENT**

We acknowledge the provision of high-quality dielectric layers by Damon Farmer (IBM Research, USA), oil emulsion by Johann Penuela (PUC, Rio de Janeiro, Brazil), and expert technical assistance by Bruce Ek (IBM Research, USA). Furthermore, we acknowledge support by Ulisses Mello (IBM Research, Brazil) and Shu-Jen Han (IBM Research, USA), as well as discussion with Ado Jorio, Luiz Gustavo Cançado, Cassiano Rabelo, and Laura Amorim (all UFMG, Belo Horizonte, Brazil).

**FIGURES**

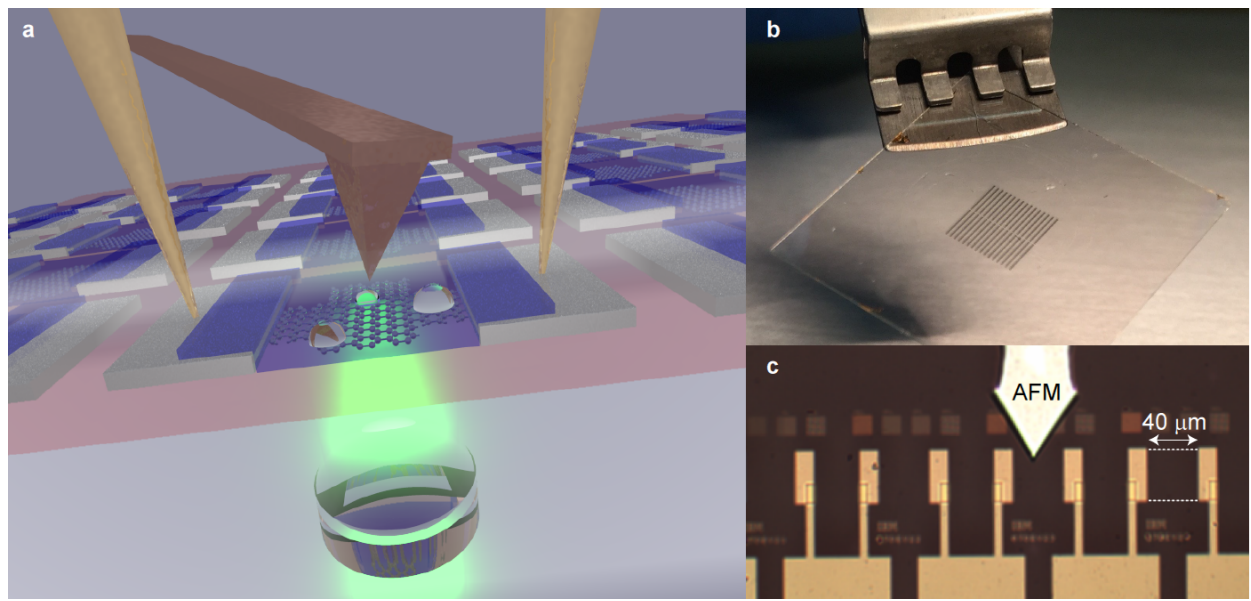

**Figure 1.** (a) Conceptual rendering of the analysis platform with integrated 2-d sensor array for simultaneous optical, electronic, and topographic measurements performed on nanoscale liquid droplets. (b) A photograph of the optically transparent analysis chip exhibits the integrated 2-dimensional device array in the platform center. (c) Optical micrograph of the platform showing individual measurement sites within the integrated 2-d sensor array. An atomic force microscope (AFM) cantilever is approaching the platform from the top. The position and active area of an individual 2-d sensor site is highlighted by dashed lines.

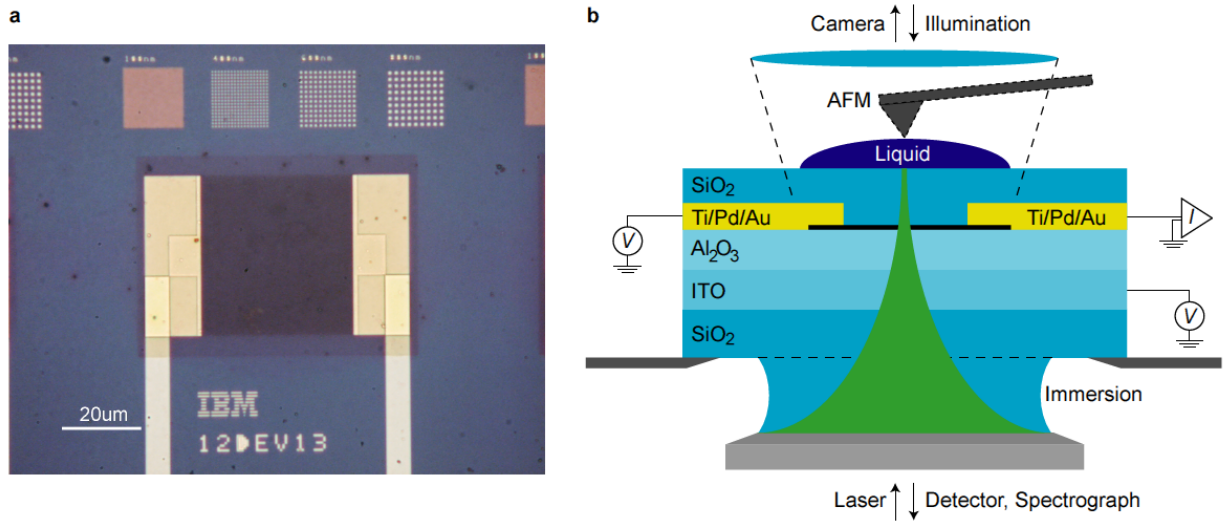

**Figure 2.** (a) Individual measurement site of the analysis platform with integrated 2-d sensor. (b) Schematic cross section of the analysis platform with integrated 2-d sensor indicating optical, electronic, and topographic measurement modes. The device stack of the platform is designed to be operated within an inverted immersion microscope for performing confocal micro-spectroscopy from the underside of the sample with high optical excitation and collection efficiency. Each device can be addressed individually and connected to an external electric transport measurement system. From the top of the analysis platform, atomic force microscopy as well as optical microscopy can be applied.

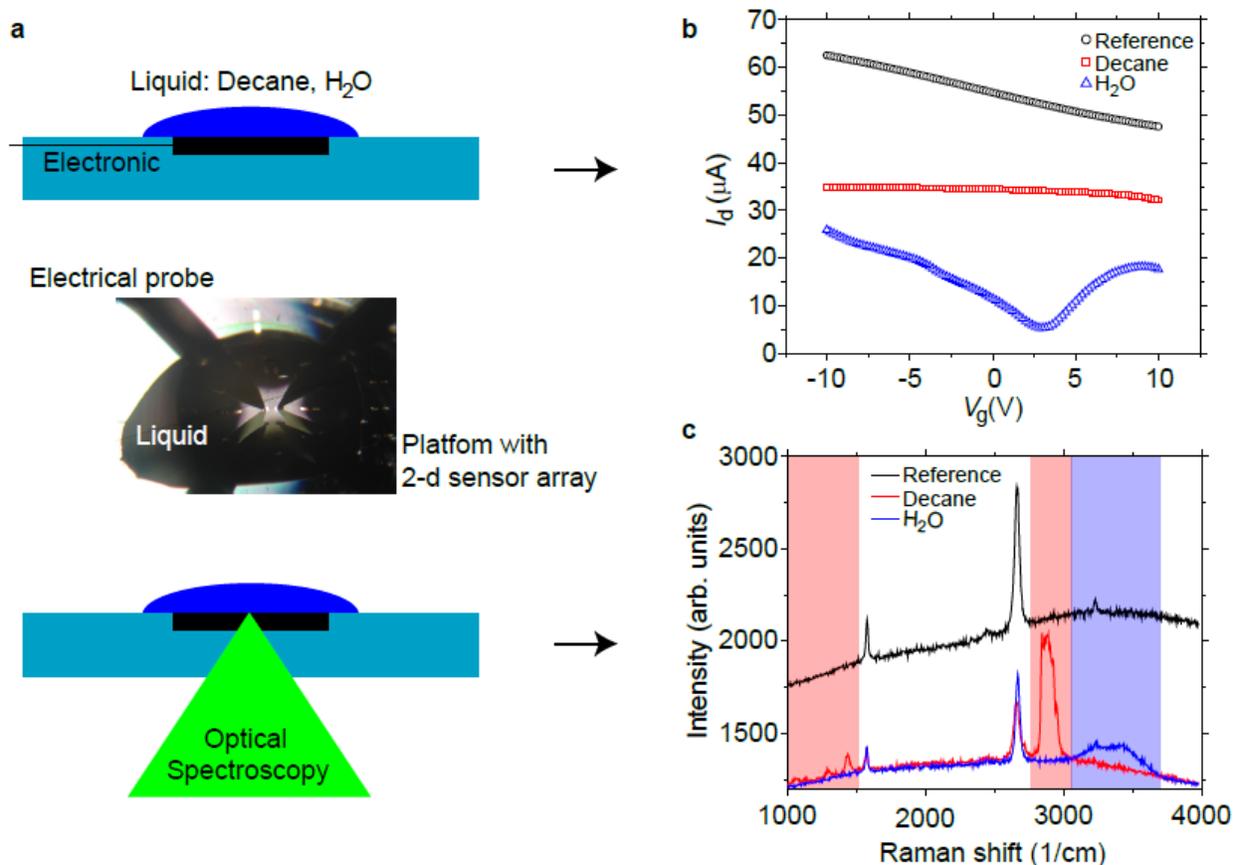

**Figure 3.** (a) Visualization of the measurement principle for integrated electronic sensing (top) and confocal optical spectroscopy (bottom) of liquids at the position of an integrated 2-d sensor within the analysis platform. The photograph (middle) shows a liquid droplet covering an integrated device site in measurement configuration. In (b), electric transfer characteristics are shown that were obtained in case of deposition of water and decane, respectively. The transfer characteristics of the integrated graphene device in air is also shown as a reference. In (c), the micro-Raman spectra acquired from the underside of the platform reveal the chemical identity of the two liquids under investigation. The shaded colored areas highlight the spectral regimes where vibrational modes of decane (red) and water (blue), respectively, are observable. The reference spectrum taken without liquid exhibits the principal Raman bands of graphene.

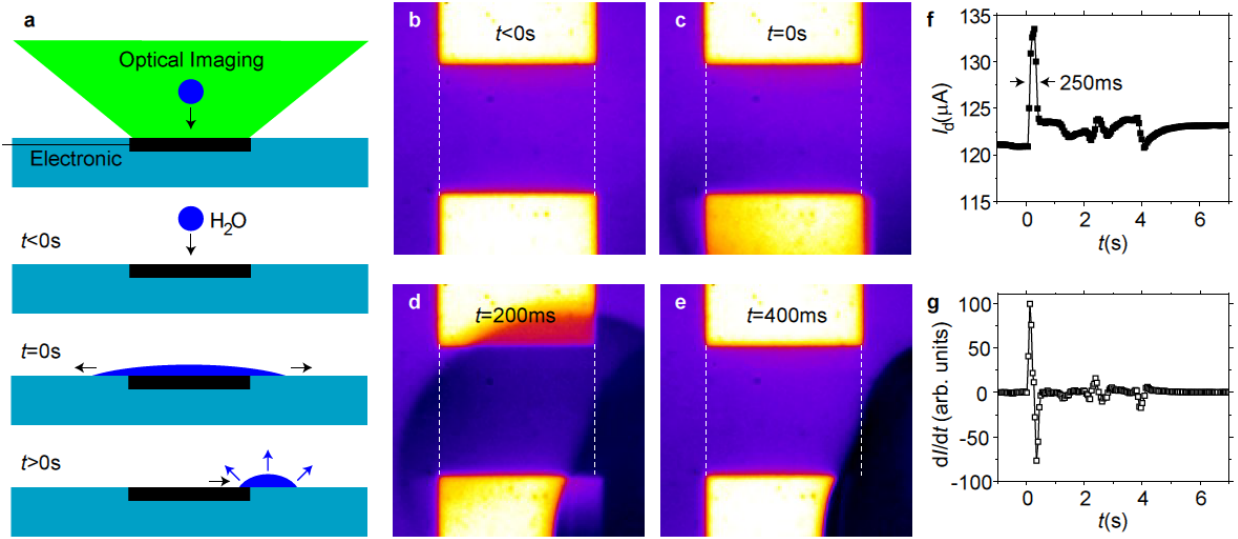

**Figure 4.** (a) Visualization of the measurement method for simultaneous optical wide field imaging and electronic sensing of wetting dynamics with the integrated 2-d sensor. Droplet deposition, wetting and evaporation dynamics are indicated for various times $t$. (b-e) Series of false color wide field optical microscopy images taken from the top, revealing the position of the graphene layer (marked by dashed lines) and a water droplet at representative time steps. The landing water droplet becomes visible as a dark shadow in the lower half of (c), spreads across the graphene area in (d), and retracts to the lower right side of the device area in (e) where it, ultimately, evaporates. The gap between the two metallic contacts, i.e. the length of white dashed lines, is 20 micrometer. The time increment of the optical image series is $\Delta t_{opt}$=200ms. (f) Device current $I_d$ through the graphene layer as function of time. The bias voltage is $V_d$=0.1mV; the increment of the electric current time series is $\Delta t_{el}$=50ms. The full-width-at-half-maximum value of the current peak that occurs at the time of droplet deposition ($t$=0s) is indicated by arrows and is due to the water droplet spreading across the graphene layer, see (d). (g) Derivative of the device current shown in (f) as function of time. The current derivative exhibits a flat baseline for as long as there is no liquid on the active sensor area.

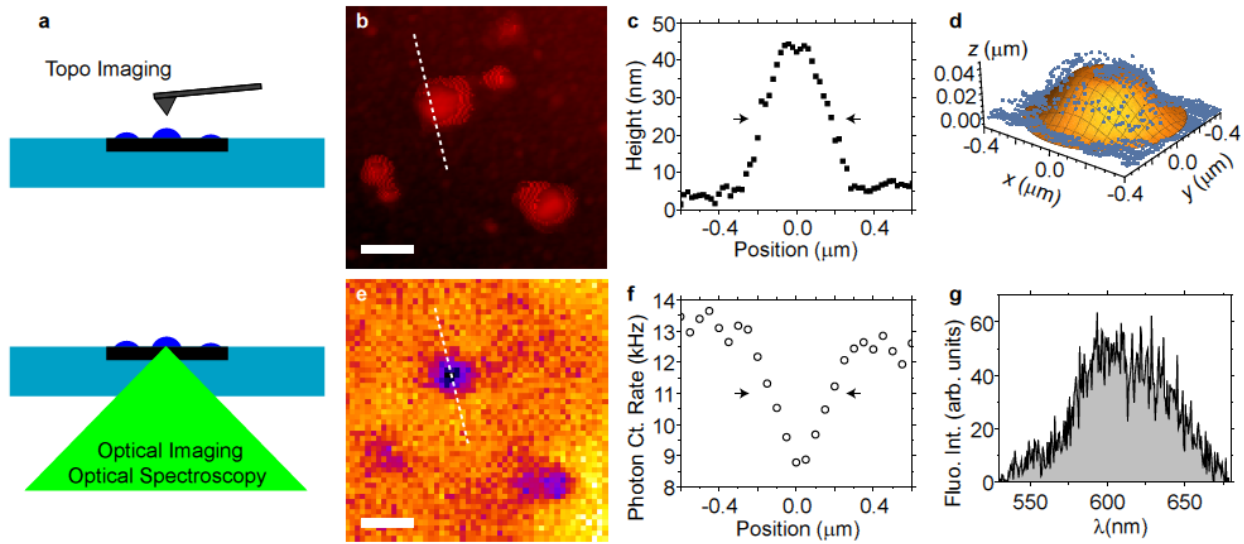

**Figure 5.** (a) Visualization of the measurement methods for nanoscale droplet characterization; atomic force microscopy from the top and confocal micro-spectroscopy from the underside of the platform at the location an integrated 2-d sensor. (b) Atomic force microscope image of individual, spatially isolated nanoscale oil emulsion droplets at the device surface. Length of white scale bar: 500nm; false color intensity scale: black (-36nm) – red (30nm). (c) Topographical cross section of an oil emulsion droplet along the white line in (b) exhibits a full-with-at-half maximum value of 350nm as indicated by arrows. (d) Three dimensional fit (surface) to the experimental topography data (dots) for the same droplet. (e) Confocal laser elastic scattering microscope image of the same device area imaged in (b) exhibits negative optical contrast at the position of nanoscale oil emulsion droplets. Length of white scale bar: 500nm; false color intensity scale: black (8.52kHz) – white (15.86kHz). (f) Light scattering intensity cross section taken along the dashed line in (e) reveals the width of the droplet contact area at the device surface with full-with-at-half maximum value of 350nm which is indicated by arrows. (g) Fluorescence spectrum of the nanoscale oil emulsion droplet marked by white lines in (b), (e).

20